# Multifunctional polymer nanofibers: UV emission, optical gain, anisotropic wetting and high hydrophobicity for next flexible excitation sources

Giovanni Morello[a,b], Rita Manco[a,c], Maria Moffa[a], Luana Persano[a], Andrea Camposeo[a,b], and Dario Pisignano[a,b,c,*]

[a] Istituto Nanoscienze-CNR, Euromediterranean Center for Nanomaterial Modelling and Technology (ECMT), via Arnesano, Lecce I-73100, Italy

[b] Center for Biomolecular Nanotechnologies @UNILE, Istituto Italiano di Tecnologia, Via Barsanti, I-73010 Arnesano (LE), Italy

[c] Dipartimento di Matematica e Fisica "Ennio De Giorgi", Università del Salento, via Arnesano I-73100 Lecce, Italy

*Corresponding author: dario.pisignano@unisalento.it

ABSTRACT

The use of UV light sources is highly relevant in many fields of science, being directly related to all those detection and diagnosis procedures which are based on fluorescence spectroscopy. Depending on the specific application, UV light-emitting materials are desired to feature a number of opto-mechanical properties, including brightness, optical gain for being used in laser devices, flexibility to conform with different lab-on-chip architectures, and tailorable wettability to control and minimize their interaction with ambient humidity and fluids. In this work, we introduce multifunctional, UV-emitting electrospun fibers with both optical gain and anisotropic hydrophobicity greatly enhanced compared to films. Fibers are described by the onset of a composite wetting state and their arrangement in uniaxial arrays further favours liquid directional control. The low gain threshold and optical losses, the plastic nature and the flexibility and stability of these UV-emitting fibers make them interesting for building light-emitting devices and microlasers. Furthermore, the found anisotropic hydrophobicity is strongly synergic with optical properties, reducing interfacial interactions with liquids and enabling smart functional surfaces for droplet microfluidic and wearable applications.

INTRODUCTION

A lot of modern research on materials is in the wide field of human healthcare and safety, involving significant issues such as the protection from chemical and biological hazard, the detection of toxic combustions, and the development of novel, low-cost point-of-care diagnostics. In these areas, the use of UV light sources is highly relevant, being related to all those detection and diagnosis methods which are based on fluorescence spectroscopy.[1,2] Depending on the specific application, UV light-emitting materials are desired to feature a number of opto-mechanical properties. These include bright emission, stimulated emission and optical gain for being used in laser devices, flexibility to conform with different lab-on-chip architectures, and tailorable wettability to control and minimize the interaction of emitting materials with ambient humidity and fluids.[3]

Inorganic materials and nanostructures have received a lot of attention for realizing UV-light emitting devices (e.g., in plasmonic nanolasers).[4-8] These materials can be processed by colloidal methods, controlled growth, or top-down lithographies. However, all of them suffer from some drawbacks, such as the poor flexibility, the relatively high production cost, and the frequently harsh processing conditions. Organic nanomaterials, on the other hand, are inherently soft and can be processed at low cost, which might greatly help in realizing flexible light sources. For instance, various UV light-emitting devices and lasers have been produced by spin-cast films and organic slabs,[9-12] while soft nanomaterials with emission and optical gain in the UV are still largely unexplored.[13]

Nowadays, a great progress in the structural flexibility and in the versatility of the surface properties of photonic materials is being offered by electrospun polymer nanowires.[14-20] These fibers show high optical anisotropy and internal molecular orientation,[21-24] enhanced quantum yield,[25] and eventually optical gain, which has led to demonstrate lasing in the visible[26,27] and near-infrared[28] spectral range by individual nanowires or mats,[26,28] and by distributed feedback imprinted geometries.[27] However, electrospun nanofibers approaching or emitting in the UV spectral range have only been based on hybrid structures embedding ZnO or Ge nanocrystals, whose fabrication involves additional steps of thermal annealing, laser ablation, or chemical reactions.[29,30] The development of purely organic, flexible UV-emitting fibers enabling multiple functions, hopefully including optical gain and tailored wetting, would open new perspectives for the integration of light-emitting materials in point-of-care diagnostics and sensors, and for the implementation of novel device platforms such as flexible microlasers and smart wearable fabrics. In this work, we introduce multifunctional, UV-emitting electrospun fibers with both optical gain and anisotropic hydrophobicity. Hydrophobic properties help in making optical properties stable, reducing interfacial interactions of active materials with water, and make these fibers good candidates for integration in lab-on-chip and in droplet microfluidic platforms for chromophore excitation in liquids. In particular, line narrowing is found for the first time in electrospun fibers emitting in the UV, with onset for excitation fluence of a few tens of 50 $\mu J/cm^2$ and optical gain above 5 $cm^{-1}$. These features, together with the flexibility and conformability of the fiber mats, suggest applications in various light-emitting device and microlaser architectures with controllable and complex shape.

## RESULTS AND DISCUSSION

Fibers embed the molecular dye, 4,4'''-bis[(2-butyloctyl)oxy]-1,1':4',1'':4'',1'''-quaterphenyl (BBQ, molecular structure in Figure 1a), and their bundles are highly effective excitation sources for other visible-emitting chromophores and labelled antibodies.

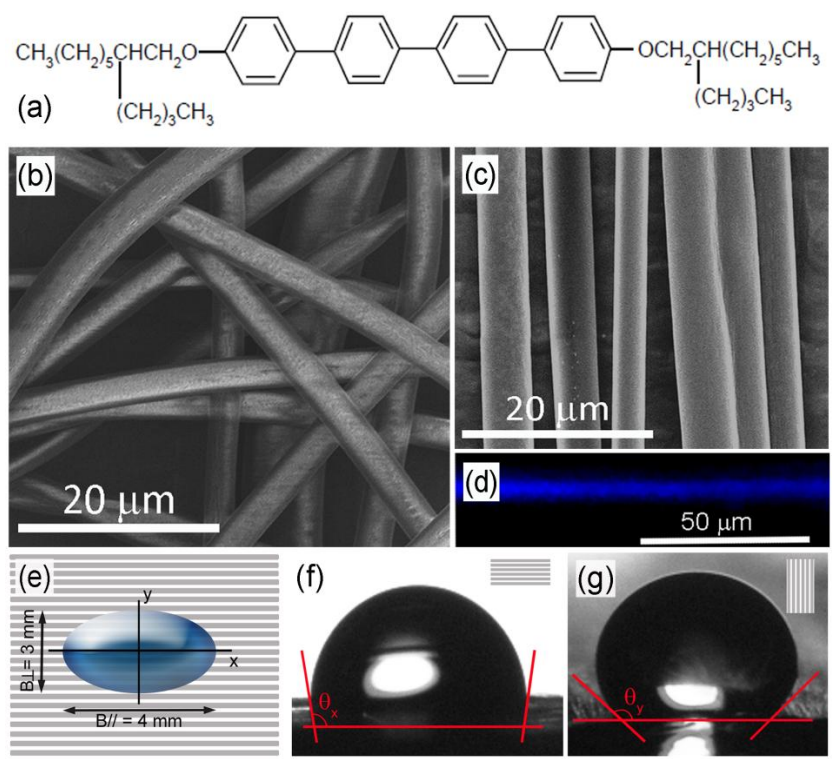

**Figure 1.** (a) BBQ molecular structure. (b) SEM micrograph of BBQ-doped electrospun fibers with random orientation. The fibers are uniform, defect-free and highly flexible. (c) Unidirectional, aligned fibers. (d) Fluorescence micrograph of a single fiber, showing a uniform and bright emission and highlighting the homogeneous presence of the embedded dye and the lack of appreciable clustering effects. (e-g) Schematics (e) and optical micrographs of water droplets and corresponding WCA values for the direction parallel (*x*-direction, f) and perpendicular (*y*-direction, g) to the fiber alignment axis.

The absorption, photoluminescence, and amplified spontaneous emission (ASE) spectra of BBQ are displayed in Figure S1 in the Supporting Information. Electrospinning allows obtaining defect-free, BBQ-doped fibers, with uniform internal distribution of the light-emitting component (Figure 1b-d). Fibers can be easily deposited in compact, free-standing bundles made by arrays of aligned wires (Figure 1c), thus favouring the assessment of optical gain properties, which have been found to be affected by the mutual alignment of electrospun fibers.[28] The topography of UV-emitting fiber surfaces also strongly affects their wettability. Indeed, while BBQ-doped poly(methylmetacrylate) (PMMA) films are hydrophilic, exhibiting a water contact angle (WCA), $\theta=61°\pm2°$ (Figure S2a), randomly-oriented fibers are remarkably hydrophobic, with WCA, $\theta_{F,Rand}=132°\pm2°$ (Figure S2b), not varying significantly upon changing the doping level in the range 0.1-2% wt/wt. Hence, the fiber topography leads to the onset of a composite wetting state described by the Cassie-Baxter law, $\cos(\theta_{F,Rand})=f_{CB}(\cos\theta+1)-1$, where $f_{CB}<1$ is the fraction of the solid surface wet by the liquid.[31] This regime is related to the presence of air trapped within the surface of neighbor electrospun fibers. In addition, while on randomly-oriented fibers wetting is isotropic and the deposited droplets take a spherical shape, aligning fibers in uniaxial arrays leads to an asymmetric surface topography and consequently to different WCA measured along a direction parallel (//) or perpendicular ($\perp$) to the fiber longitudinal axis, respectively. Pinning on the fibers tightly confines the fluid, and induces an asymmetric shape of deposited water droplets (scheme in Figure 1e).[32-34] A stronger confinement is found perpendicular to the fiber length, with WCA, $\theta_{F,Align}^{//}=99°\pm2°$ (Figure 1f) and $\theta_{F,Align}^{\perp}=140°\pm2°$ (Figure 1g), approaching a superhydrophobic regime.[35] This result is consistent with the Gibbs' criterion for liquid pinning, according to which the wetting contact line remains pinned on a microstructure when the advancing contact angle is comprised between $\theta$ and $\theta+\alpha$, where $\alpha$ ($\simeq 90°$) is the maximum inclination of the surface of individual deposited fibers with respect to their top edge.[34] At equilibrium, the shape of drops on the fiber surface can be predicted by minimizing the reduced free energy, $f=G/\gamma_{LV}$, where $G$ is the overall free energy and $\gamma_{LV}$ indicates the liquid-vapor surface tension.[36-38] $f$ in the direction parallel and perpendicular to the fiber longitudinal axis can be estimated by[37] $f_{//}= [B_{//}\theta_{F,Align}^{//}/ \sin (\theta_{F,Align}^{//})]- B_{//}\cos\theta$ and $f_{\perp} = [B_{\perp}\theta_{F,Align}^{\perp}/ \sin (\theta_{F,Align}^{\perp})]- B_{\perp}\cos\theta$, where $B_{\perp}$ and $B_{//}$ are the lengths of the drop base in the two directions. For droplets of $B_\perp \cong 3$ mm and $B_{//} \cong 4$ mm (Figure 1e), we find a difference of about 40% for the corresponding reduced free energies. Overall, electrospun fibers show greatly enhanced hydrophobicity compared to films, and their arrangement in uniaxial arrays further favors liquid directional control through pinning effects. Both these effects might contribute to reduce interfacial interactions of the light-emitting material with water, thus improving both mechanical and optical stability. In fact, wetting properties are also related to the fiber surface chemistry and polarity, and variations of the surface polarity, like those associated to oxidation processes, can be unravelled by WCA changes.[39] The UV-emitting fibers, excited under conditions both below and above the ASE threshold (50-300 $\mu J/cm^2$, see the following) do not show appreciable wettability variations, thus evidencing excellent surface chemical stability.

UV fiber bundles are then optically characterized with the aim of assessing ASE threshold, net gain and optical losses. In Figure 2 we report the results of ASE measurements, showing emission spectra obtained at different excitation fluences, and the resulting behavior of the spectral linewidth and emission intensity for fibers. At a low excitation level (<50 $\mu J/cm^2$) the signal keeps low and the spectra are broad (see also Figure S3), mainly consisting of spontaneous emission.

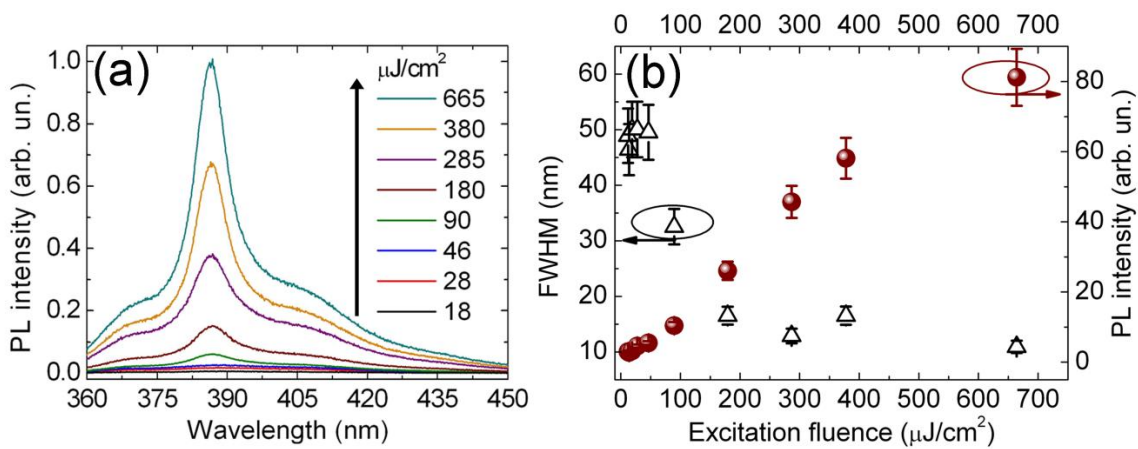

**Figure 2.** (a) ASE spectra from BBQ-doped fibers for various excitation fluences. (b) ASE intensity (full symbols) and FWHM (empty symbols) *vs.* excitation fluence.

At about 50 µJ/cm$^2$ the spectra of fibers start to narrow, peaking around 387 nm. Spectral narrowing is an effect typically related to ASE, which is induced by photons travelling across the length of optical gain in an active medium.[40] In materials exhibiting stimulated emission, under intense pump excitation line-narrowing occurs in the spectral regions where the gain, which is related to the peak cross-section of the involved optical transition, is higher. The effect is frequently assisted by photon waveguiding along the gain material, which may constitute the core of an asymmetric waveguide in which the organics are sandwiched between two media with lower refractive index (the quartz substrate underneath and vacuum). The here found peak wavelength well-matches the spectral region of state-of-art inorganic microlasers such as those based on ZnO nanowires.[6,7] Up to 300 µJ/cm$^2$ the ASE intensity grows in a superlinear way, whereas the full width at half maximum (FWHM) undergoes an abrupt decrease down to 11 nm. Spectral narrowing mainly occurs around 100 µJ/cm$^2$. Overall, fibers show good ASE performances, which is particularly promising in view of building nanofiber-based, flexible UV-

Published in ACS Applied Materials & Interfaces: 7, 21907–21912. doi: [10.1021/acsami.5b06483](10.1021/acsami.5b06483) (2015).microlasers. The net gain $G(\lambda)$ at 387 nm, shown in Figure 3a, is then obtained from the following expression:

$$I_L = \frac{I_p A(\lambda)}{G(\lambda)} \cdot \left[e^{G(\lambda) \cdot L} - 1\right], \tag{1}$$

where $I_L$ is the emission output intensity, $I_p$ represents the pump intensity, $A(\lambda)$ is related to the spontaneous emission cross-section, and $L$ is the length of the excited region (corresponding film data are shown in Figures S4 and S5). As expected above threshold, fibers show an exponential raise of the output intensity upon varying the stripe length (inset of Figure 3a), whereas the gain spectrum resembles the spectral profile of ASE. A maximum gain of 5.4 cm$^{-1}$ is measured at 387 nm, which is comparable with values of bright, dye-doped polymer fibers emitting in the visible and near-infrared range.[28] The slightly higher value measured for films (Figure S5a) can be rationalized by considering that the estimated $G$ is given by the difference of the optical gain coefficient in the active material and the optical loss coefficient, $\gamma$. Optical losses are typically higher in electrospun fibers compared to film slabs, due to the more effective light scattering in the microstructures (Fig. 3b and Fig. S5b).

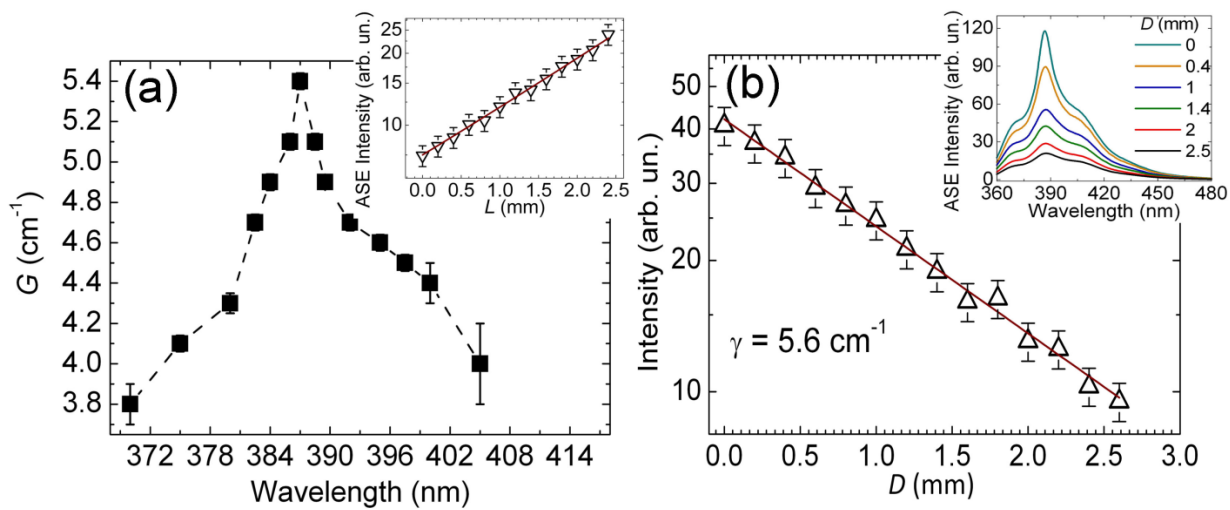

**Figure 3.** (a) Gain spectrum of BBQ-doped fibers. The plotted net gain values (symbols) are obtained by analyzing the ASE intensity as a function of the excitation stripe length at each wavelength by means of Eq. (1). The dashed line is a guide for the eyes. Inset: example of the dependence of the ASE intensity on the excitation stripe length (empty symbols) and corresponding best fit (continuous line). Data correspond to the maximum gain wavelength (387 nm). (b) Plot of the edge-emission intensity, recorded at different distances ($D$) of the excitation stripe from the fiber termination (empty symbols). The continuous line is the best fit by an exponential decay, providing the optical loss coefficient, $\gamma$. Inset: spectra collected at different $D$ values. Excitation fluence = 285 $\mu J/cm^2$.

Another important feature of light-emitting nanofibers is the attenuation occurring during the propagation of guided light. To assess optical losses, we measure the emission intensity as a function of the varying distance, $D$, of the excitation region from the sample emitting edge. The output intensity is then fitted by the expression, $I_{PL} = I_0 \cdot e^{-\gamma D}$, where $I_0$ is a constant and $\gamma$ is the loss coefficient. Figure 3b shows the resulting output intensity from the fiber sample excited at 285 $\mu J/cm^2$, providing $\gamma$ = 5.6 cm$^{-1}$. This is a very favorable value compared to other light-emitting nanofiber systems,[28] the low optical losses here being related to the reduced contribution of self-absorption of emitted light because of the large Stokes shift of BBQ (766 meV, see Figure S1).

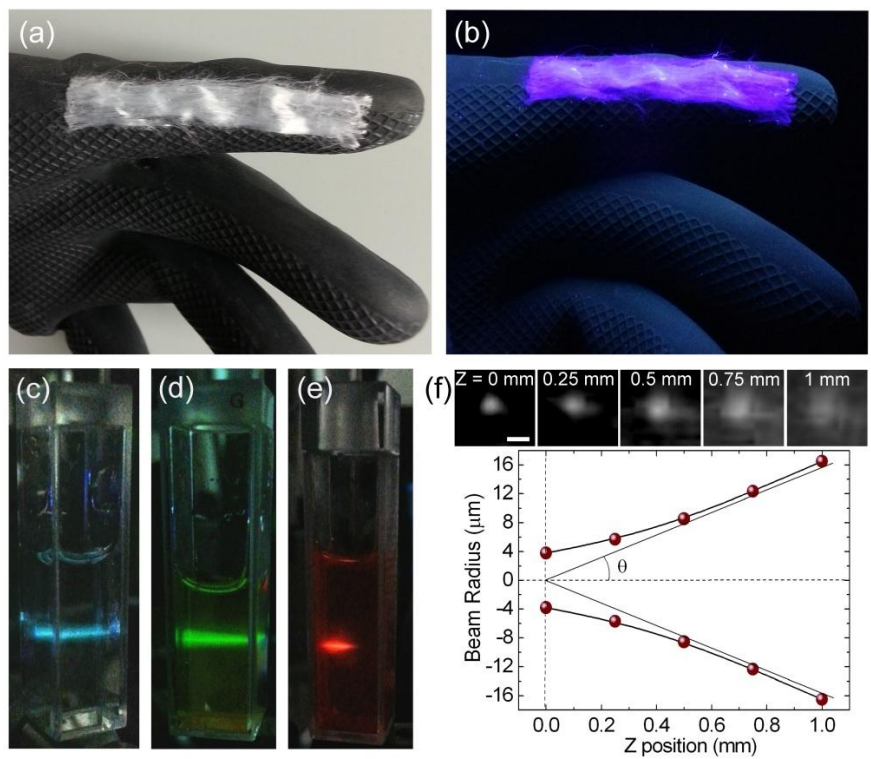

**Figure 4.** (a) Photograph of a bundle of aligned BBQ-doped fibers on curved surfaces. (b) The same bundle while emitting UV light. (c-e) Photoluminescence of 4',6-Diamidino-2-phenylindole (DAPI, blue-emitting, c), 2-(6-Amino-3-imino-3H-xanthen-9-yl)benzoic acid methyl ester (Rhodamine 123, green-emitting, d), and [2-[2-[4-(dimethylamino)phenyl]ethenyl]-6-methyl-4H- pyran-4-ylidene]-propanedinitrile (DCM, red-emitting, e) solutions, excited by the UV ASE from BBQ-doped fibers. Further experimental details for this configuration are in Fig. S6. (f) ASE beam divergence measurement. Top: micrographs of the emission from a single fiber, recorded at different positions from the emitting fiber tip image (Z position). Scale bar = 10 μm. Bottom: plot of the beam radius (dots) *vs*. Z. According to the definition of the divergence, $\theta$, of a Gaussian beam, we consider the half angle corresponding to the asymptotic variation of the beam radius along the light propagation direction.

Furthermore, UV-emitting fiber bundles are mechanically robust, can be repeatedly bent, and reversibly conform to both planar and curved surfaces (Figure 4a,b). The reversible adhesion contact to other surfaces, mediated by Van der Waals forces, allows for repeated use of the same bundles or mats with different substrates. Hence, these materials can be used as flexible excitation sources in a variety of spectroscopic measurements and lab-on-chip architectures, inducing fluorescence in chromophores and labelled antibodies with emission in the visible and in the near-infrared, under either continuous or pulsed excitation conditions. For example, in Figure 4c-e we show how the ASE signal from a bundle of aligned BBQ-based fibers excites three different dyes dissolved in solution. It is worth noting that the ASE beam profile is highly directional (Figure 4f) with an estimated divergence of 16.5 mrad, in line with the smaller divergence reported for polymers with optical gain.[41]

In particular, digital microfluidic applications can strongly benefit from the synergy of the emission and wetting properties of these fibers. To this aim, fibers aligned on a glass or quartz surface can be pumped by an evanescent field through total internal reflection (TIR)[42] and in turn used to excite fluorescent droplets of solutions of chromophores or labeled antibodies (Figure 5). Droplets are placed and moved on fibers without touching the glass surface underneath, which is favored by the liquid confinement and hydrophobic character of the bundles, making this material ideal as active substrate for fluorescence excitation in digital microfluidics.

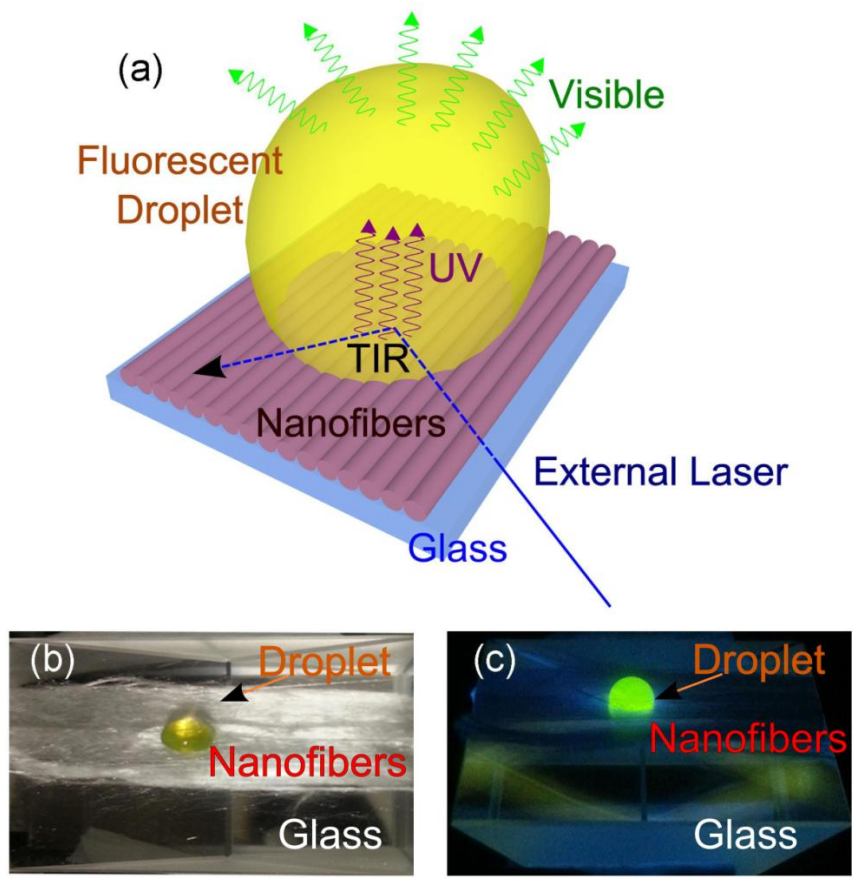

**Figure 5.** Aligned UV-light emitting nanofibers as smart substrate for droplet microfluidics. (a) Schematics of a bundle of fibers with thickness of a few μm, deposited onto a prism, and evanescently excited by laser light (355 nm) in a TIR configuration. (b, c) Bright-field (b) and fluorescence (c) pictures of a droplet of Rhodamine 123 fluorescent solution in water. In (c), the fibers UV emission excites the droplet, and it is then removed by a long-pass filter.

### CONCLUSION

In conclusion, we demonstrate multifunctional UV-emitting, BBQ-doped organic fibers, featuring ASE, optical gain, as well as anisotropic wetting and hydrophobicity. The low ASE threshold and optical losses, the plastic nature and the flexibility of these UV-emitting filaments make them interesting for building light-emitting devices and microlasers. Furthermore, the found anisotropic hydrophobicity is strongly synergic with optical properties, reducing interfacial interactions of the active material with liquids and enabling smart functional surfaces for microfluidic and wearable applications.

## MATERIALS AND METHODS

**Electrospinning.** 1% wt/wt BBQ/PMMA solutions are prepared by $CHCl_3$ (PMMA content = 3.75% wt compared to solvent). The concentration is chosen in the range 0.1-2%, following preliminary tests performed to optimize the emission. The solution is mixed by mechanical stirring for 12 h at room temperature prior electrospinning, which is performed by a 1 mL syringe and a 21 gauge stainless needle. Randomly-oriented fibers are spun by an applied voltage of 15 kV (EL60R0.6-22, Glassman High Voltage) and a needle-collector distance of 20 cm. The electrospinning injection flow rate is kept constant at 1 mL $h^{-1}$ using a syringe pump (Harvard Apparatus). Aligned fibers are obtained using a collector having a diameter of 8 cm and rotating at 4000 rpm, placed at 10 cm from the needle. Reference films are spin-cast from the same solution used for electrospinning. Sample inspection is performed by a Nova NanoSEM 450 system (FEI).

**Contact angle measurements.** The apparent WCA on the surface of films and electrospun fibers is measured by an optical video contact angle system (CAM-200, KSV Instruments), gently delivering a drop (ca. 20 µL) of ultrapure water from a capillary tip onto the fiber surface. The contact angles are determined by fitting the profile of at least three droplets. WCA measurements are performed both before and after UV pulsed light irradiation (355 nm, exposure time = 3 s).

**Optical characterization.** Absorption measurements are carried out by a spectrophotometer. Spontaneous emission spectra are obtained by exciting samples below gain threshold. Fluorescence micrographs are collected by an inverted microscope (Nikon), exciting fibers with a mercury lamp through an objective lens (50×, NA = 0.75). The fiber emission is then collected by the same objective and measured by a charge-coupled device (CCD). Optical excitation of fluorescent dyes in water is performed by directing ASE from one edge of the fibers sample into filled cuvettes. Evanescent field excitation of solution droplets deposited onto fiber mats is obtained by using bundles of aligned fibers on top of a prism. The excitation laser (355 nm) is directed into the prism, being reflected at the internal surface at an angle larger than the critical one according to a TIR regime. The fiber emission is cut by a long-pass filter after exciting the fluorescent droplets, thus being not collected by the imaging system.

**Optical gain.** Bundles of aligned fibers and spin-coated films are placed under vacuum at $10^{-4}$ mbar and excited by the third harmonic (355 nm) of a pulsed Nd:YAG laser (pulse duration = 10 ns, repetition rate = 10 Hz) focused on a stripe parallel to the fiber alignment direction. A system of lenses allows collecting the emitted light from one edge of the fibers, coupling the photoluminescence signal into a spectrograph. The ASE threshold is then measured by systematically varying the excitation fluence, and the net gain is obtained by the variable stripe length technique. By keeping fixed the excitation fluence above threshold as well as the stripe position on the sample, we change the stripe length and measure the emitted intensity, finally analyzed by Equation (1). In a different way, optical losses are measured through the emitted light intensity by varying the excitation stripe position (with fixed excitation fluence and length = 4 mm), namely moving the stripe away from the sample emitting edge. For beam divergence measurements, a bundle of aligned fibers is excited above threshold, and ASE is coupled to a lens system (numerical aperture 0.85, resolving power of about 0.3 μm). A Si CCD is initially placed on the focus plane at which the tip of the emitting fiber is imaged (Z=0 mm in Figure 4f) and gradually moved away in order to record the divergent emission profile as a function of the distance from the focus position.The plot of the so obtained spot diameter *vs.* the CCD position allows the beam divergence to be estimated, which is defined as the half angle corresponding to the asymptotic variation of the beam radius along the propagation direction.

ASSOCIATED CONTENT

**Supporting Information**. Further wettability and optical data, and experimental details. This material is available free of charge via the Internet at http://pubs.acs.org.

ACKNOWLEDGMENT

The research leading to these results has received funding from the European Research Council under the European Union's Seventh Framework Programme (FP/2007-2013)/ERC Grant Agreement n. 306357 ("NANO-JETS").

REFERENCES

(1) Misra, P.; Dubinskii, M. A. *Ultraviolet Spectroscopy and UV Lasers*, 1st ed. Marcel Dekker, Inc. New York, Basel, Switzerland, 2002.

(2) Schneider, D.; Rabe, T.; Riedl, T.; Dobbertin, T.; Kroger, M.; Becker, E.; Johannes, H-H.; Kowalsky, W.; Weimann, T.; Wang, J.; Hinze, P.; Gerhard, A.; Stossel, P.; Vestweber, H. An Ultraviolet Organic Thin-Film Solid-State Laser for Biomarker Application. *Adv. Mater.* **2005**, *17*, 31-34.

(3)  Wu, H.; Zhang, R.; Sun, Y.; Lin, D.; Sun, Z.; Pan, W.; Downs, P. Biomimetic Nanofiber Patterns with Controlled Wettability. *Soft Matter* **2008**, *4*, 2429-2433.

(4)  Pauzauskie, P. J.; Sirbuly, D. J.; Yang, P. Semiconductor Nanowire Ring Resonator Laser. *Phys. Rev. Lett.* **2006**, *96*, 143903-143906.

(5)  Liu, H.; Hu, L.; Watanabe, K.; Hu, X.; Dierre, B.; Kim, B.; Sekiguchi, T.; Fang, X. Cathodoluminescence Modulation of ZnS Nanostructures by Morphology, Doping, and Temperature. *Adv. Funct. Mater.* **2013**, *23*, 3701-3709.

(6)  Zhou, H.; Wissinger, M.; Fallert, J.; Hauschild, R.; Stelzl, F.; Klingshim, C.; Kalt, H. Ordered, Uniform-Sized ZnO Nanolaser Arrays. *Appl. Phys. Lett.* **2007**, *91*, 181112-181114.

(7)  Xu, C. X.; Sun, X. W.; Yuen, C.; Chen, B. J.; Yu, S. F.; Dong, Z. L. Ultraviolet Amplified Spontaneous Emission from Self-Organized Network of Zinc Oxide Nanofibers. *Appl. Phys. Lett.* **2005**, *86*, 011118-011120.

(8)  Zhang, Q.; Li, G.; Liu, X.; Qian, F.; Li, Y.; Sum, T. C.; Lieber, C. M.; Xiong, Q. A Room Temperature Low-Threshold Ultraviolet Plasmonic Nanolaser. *Nat. Commun.* **2014**, *5*, 4953-4961.

(9)  Qiu, C. F.; Wang, L. D.; Chen, H. Y.; Wong, M.; Kwok, H. S. Room-Temperature Ultraviolet Emission from an Organic Light-Emitting Diode. *Appl. Phys. Lett.* **2001**, *79*, 2276-2278.

(10)  Sharma, A.; Katiyar, M.; Deepak, S.; Seki, S. Polysilane Based Organic Light Emitting Diodes: Simultaneous Ultraviolet and Visible Emission *J. Appl. Phys.* **2007**, *102*, 084506-084512.

(11) Spehr, T.; Siebert, A.; Fuhrmann-Lieker, T.; Salbeck, J.; Rabe, T.; Riedl, T.; Kowalsky, W.; Wang, J.; Weimann, T.; Hinze, P. Organic Solid-State Ultraviolet-Laser Based on Spiro-Terphenyl. *Appl. Phys. Lett.* **2005**, *87*, 161103-161105.

(12) Forget, S.; Rabbani-Haghighi, H.; Diffalah, N.; Siove, A.; Chénais, S. Tunable Ultraviolet Vertically-Emitting Organic Laser. *Appl. Phys. Lett.* **2011**, *98*, 131102-131104.

(13) Cui, Q. H.; Zhao, Y. S.; Yao, J. Controlled Synthesis of Organic Nanophotonic Materials with Specific Structures and Compositions. *Adv. Mater.* **2014**, *26*, 6852-6870.

(14) Pisignano, D. *Polymer Nanofibers,* Royal Society of Chemistry, Cambridge, UK, 2013.

(15) Camposeo, A.; Persano, L.; Pisignano, D. Light-Emitting Electrospun Nanofibers for Nanophotonics and Optoelectronics. *Macromol. Mater. Eng.* **2013**, *298*, 487-503.

(16) Li, D.; Xia, Y. Electrospinning of Nanofibers: Reinventing the Wheel? *Adv. Mater.* **2004**, *16*, 1151-1170.

(17) Anzenbacher, P.; Palacios, M. A. Polymer Nanofibre Junctions of Attolitre Volume Serve as Zeptomole-Scale Chemical Reactors. *Nat. Chem.* **2009**, *1*, 80-86.

(18) Agarwal, S.; Greiner, A.; Wendorff, J. H. Functional Materials by Electrospinning of Polymers. *Prog. Polym. Sci.* **2013**, *38*, 963-991.

(19) Sun, B.; Long, Y. Z.; Zhang, H. D.; Li, M. M.; Duvail, J. L.; Jiang, X. Y.; Yin, H. L. Advances in Three-Dimensional Nanofibrous Macrostructures via Electrospinning. *Prog. Polym. Sci.* **2014**, *39*, 862-890.

(20) Persano, L.; Camposeo, A.; Pisignano, D. Active Polymer Nanofibers for Photonics, Electronics, Energy Generation and Micromechanics. *Prog. Polym. Sci.* **2015**, *43*, 48-95.

(21) Kakade, M. V.; Givens, S.; Gardner, K.; Lee, K. H.; Chase, D. B.; Rabolt, J. F. Electric Field Induced Orientation of Polymer Chains in Macroscopically Aligned Electrospun Polymer Nanofibers. *J. Am. Chem. Soc.* **2007**, *129*, 2777-2782.

(22) Bellan, L. M.; Craighead, H. G. Molecular Orientation in Individual Electrospun Nanofibers Measured via Polarized Raman Spectroscopy. *Polymer* **2008**, *49*, 3125-3129.

(23) Pagliara, S.; Vitiello, M. S.; Camposeo, A.; Polini, A.; Cingolani, R.; Scamarcio, G.; Pisignano, D. Optical Anisotropy in Single Light-Emitting Polymer Nanofibers. *J. Phys. Chem. C* **2011**, *115*, 20399-20405.

(24) Richard-Lacroix, M.; Pellerin, C. Orientation and Structure of Single Electrospun Nanofibers of Poly(ethylene terephthalate) by Confocal Raman Spectroscopy. *Macromolecules.* **2012**, *45*, 1946-1953.

(25) Morello, G.; Polini, A.; Girardo, S.; Camposeo, A.; Pisignano, D. Enhanced Emission Efficiency in Electrospun Polyfluorene Copolymer Fibers. *Appl. Phys. Lett.* **2013**, *102*, 211911-211915.

(26) Camposeo, A.; Di Benedetto, F.; Stabile, R.; Neves, A. A. R.; Cingolani, R.; Pisignano, D. Laser Emission from Electrospun Polymer Nanofibers. *Small* **2009**, *5*, 562-566.

(27) Persano, L.; Camposeo, A.; Del Carro, P.; Fasano, V.; Moffa, M.; D'Agostino, S.; Pisignano, D. Distributed Feedback Imprinted Electrospun Fiber Lasers. *Adv. Mater.* **2014**, *26*, 6542-6547.

(28) Morello, G.; Moffa, M.; Girardo, S.; Camposeo, A.; Pisignano, D. Optical Gain in the Near Infrared by Light-Emitting Electrospun Fibers. *Adv. Funct. Mater.* **2014**, *24*, 5225-5231.

Published in ACS Applied Materials & Interfaces: 7, 21907–21912. doi: [10.1021/acsami.5b06483](10.1021/acsami.5b06483) (2015).(29) Zhang, J.; Wen, B.; Wang, F.; Ding, Y.; Zhang, S.; Yang, M. In Situ Synthesis of ZnO Nanocrystal/PET Hybrid Nanofibers via Electrospinning. *J. Polym. Sci., Part B: Polym. Phys.* **2011**, *49*, 1360-1368.

(30) Ortac, B.; Kayaci, F.; Vural, H. A.; Deniz, A. E.; Uyar, T. Photoluminescent Electrospun Polymeric Nanofibers Incorporating Germanium Nanocrystals. *React. Funct. Polym.* **2013**, *73*, 1262-1267.

(31) Li, X.-M.; Reinhoudt, D.; Crego-Calama, M. What Do We Need for a Superhydrophobic Surface? A Review on the Recent Progress in the Preparation of Superhydrophobic Surfaces. *Chem. Soc. Rev*. **2007**, *36*, 1350-1368.

(32) Chen, Y.; He, B.; Lee, J.; Patankar, N. A. Anisotropy in the Wetting of Rough Surfaces. *J. Colloid Interface Sci*. **2005**, *281*, 458-464.

(33) Semprebon, C.; Mistura, G.; Orlandini, E.; Bissacco, G.; Segato, A.; Yeomans, J. M. Anisotropy of Water Droplets on Single Rectangular Posts. *Langmuir* **2009**, *25*, 5619-5625.

(34) Kusumaatmaja, H.; Vrancken, R. J.; Bastiaansen, C. W. M.; Yeomans, J. M. Anisotropic Drop Morphologies on Corrugated Surfaces. *Langmuir* **2008**, *24*, 7299-7308.

(35) Zorba, V.; Stratakis, E.; Barberoglou, M.; Spanakis, E.; Tzanetakis, P.; Anastasiadis, S. H.; Fotakis, C. Biomimetic Artificial Surfaces Quantitatively Reproduce the Water Repellency of a Lotus Leaf. *Adv. Mater.* **2008**, *20*, 4049–4054.

(36) Kusumaatmaja, H.; Yeomans, J. M. Modeling Contact Angle Hysteresis on Chemically Patterned and Superhydrophobic Surfaces. *Langmuir* **2007**, *23*, 6019-6032.

(37) Zhang, D.; Chen, F.; Yang, Q.; Siac, J.; Hou, X. Mutual Wetting Transition between Isotropic and Anisotropic on Directional Structures Fabricated by Femtosecond Laser. *Soft Matter* **2011**, *7*, 8337-8342.20

(38) Zhang, X.; Mi, Y. Dynamics of a Stick−Jump Contact Line of Water Drops on a Strip Surface. *Langmuir* **2009**, *25*, 3212-3218.

(39) Wagner, N.; Theato, P. Light-induced Wettability Changes on Polymer Surfaces. *Polymer* **2014**, *55*, 3436-3453.

(40) Svelto, O. *Principles of Lasers*, 4th ed; Springer: New York, USA, 1998.

(41) Costela, A.; García, O.; Cerdán, L.; García-Moreno, I.; Sastre, R. Amplified Spontaneous Emission and Optical Gain Measurements from Pyrromethene 567-Doped Polymer Waveguides and Quasi-Waveguides. *Opt. Express* **2008**, *16*, 7023-7036.

(42) Axelrod, D. Evanescent Excitation and Emission in Fluorescence Microscopy. *Biophys. J.* **2013**, *104*, 1401-1409.

Supporting Information

# Multifunctional polymer nanofibers: UV emission, optical gain, anisotropic wetting and high hydrophobicity for next flexible excitation sources

*Giovanni Morello[a,b], Rita Manco[a,c], Maria Moffa[a], Luana Persano[a], Andrea Camposeo[a,b], and Dario Pisignano[a,b,c],\**

[a] Istituto Nanoscienze-CNR, Euromediterranean Center for Nanomaterial Modelling and Technology (ECMT), via Arnesano, Lecce I-73100, Italy

[b] Center for Biomolecular Nanotechnologies @UNILE, Istituto Italiano di Tecnologia, Via Barsanti, I-73010 Arnesano (LE), Italy

[c] Dipartimento di Matematica e Fisica "Ennio De Giorgi", Università del Salento, via Arnesano I-73100 Lecce, Italy

*Corresponding author: dario.pisignano@unisalento.it

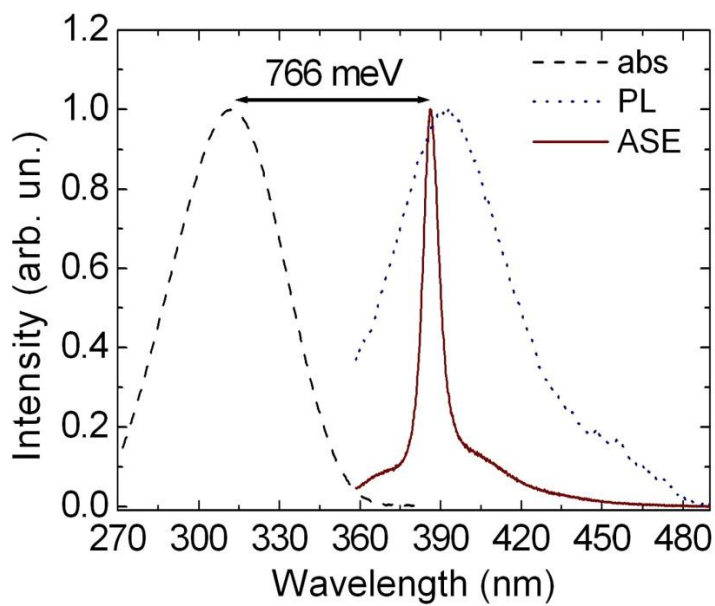

**Figure S1**. Normalized absorption (dashed line), photoluminescence (PL, dotted line) and amplified spontaneous emission (ASE, continuous line) spectra of BBQ-based fibers. The large Stokes shift (766 meV) is also highlighted.

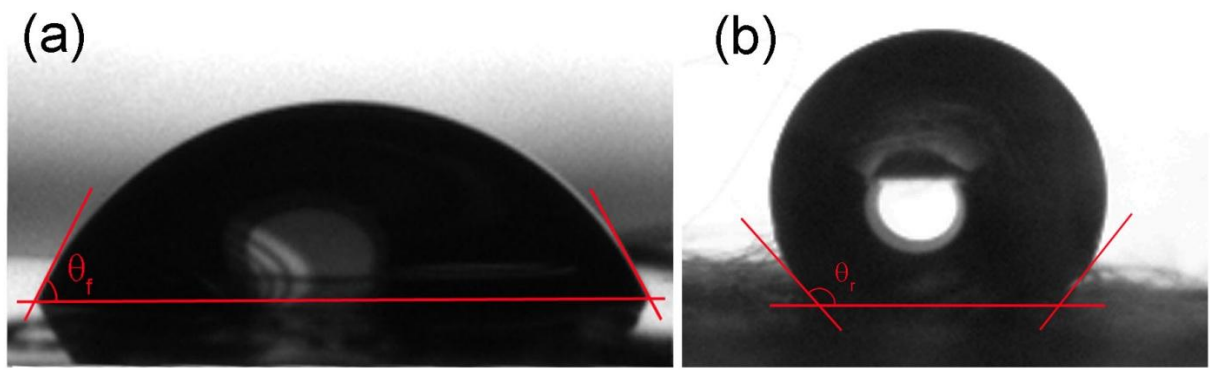

**Figure S2**. WCA on a spin-cast film (a, $\theta = 61°\pm2°$) and on a mat of randomly-oriented fibers (b, $\theta_{F,Rand}=132°\pm2°$).

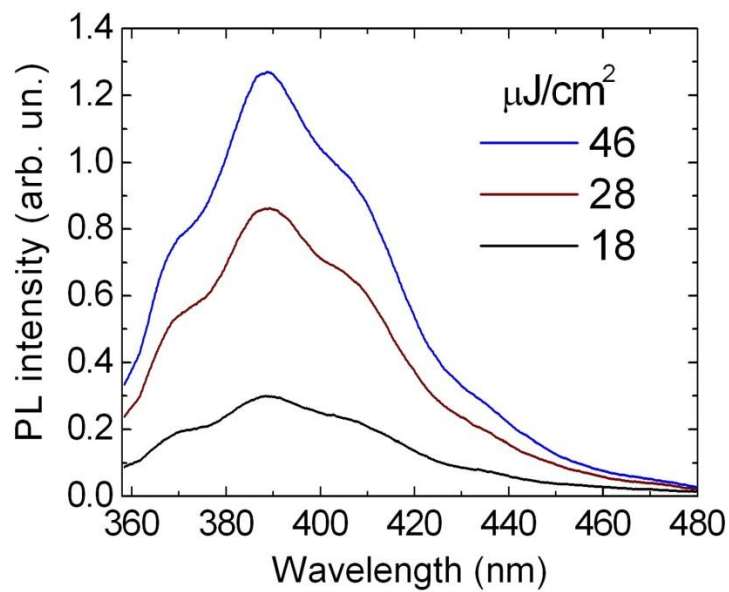

**Figure S3**. PL spectra of BBQ-doped fibers at different excitation fluences, below ASE threshold.

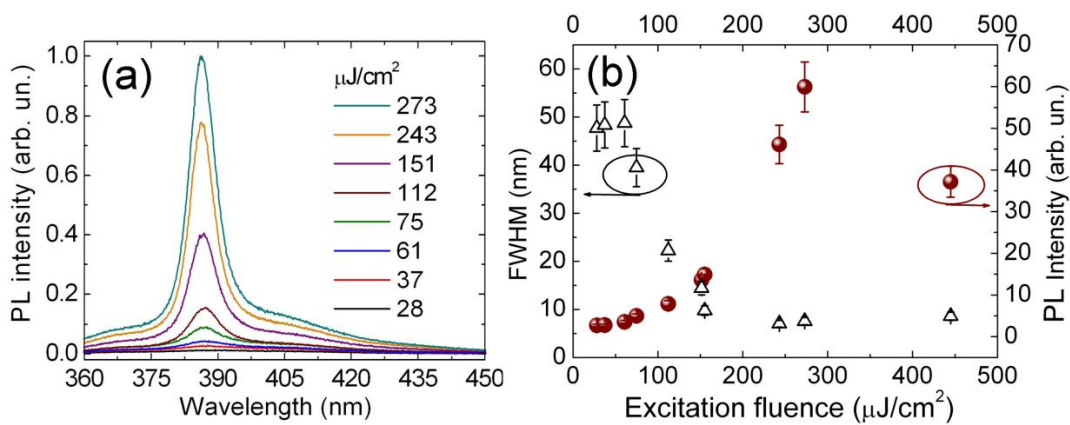

**Figure S4.** (a) ASE spectra from BBQ-doped PMMA film for various excitation fluences. (b) ASE intensity (full symbols) and FWHM (empty symbols) vs. excitation fluence.

Published in ACS Applied Materials & Interfaces: 7, 21907–21912. doi: 10.1021/acsami.5b06483 (2015).

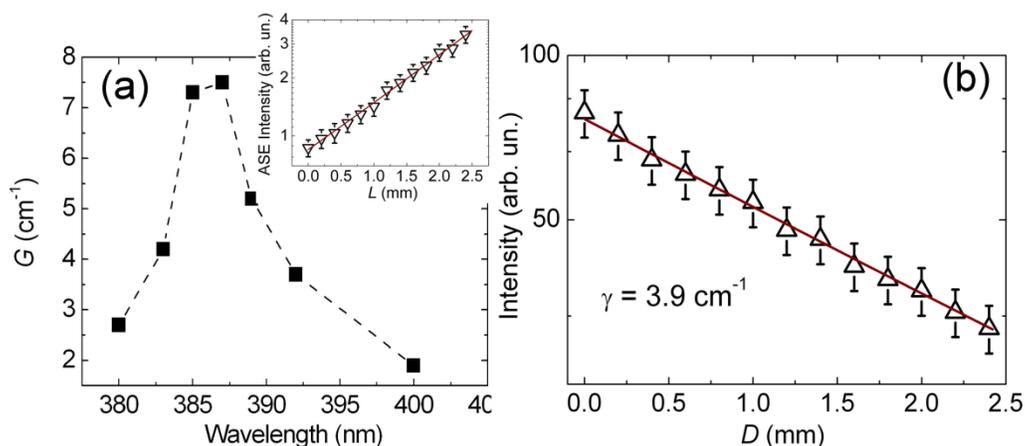

**Figure S5.** (a) Gain spectrum of BBQ-doped spin-cast PMMA films (symbols). The dashed line is a guide for the eyes. Inset: Emission intensity (symbols) vs. excitation stripe length ($L$) and corresponding best fit to Equation (1) in the main text (continuous line). (b) Output intensity vs. distance ($D$) of the excitation stripe from the emitting edge, and corresponding best fit (continuous line) by an exponential decay (see the main text). A value of $\gamma = 3.9$ cm$^{-1}$ is obtained. Excitation fluence = 150 µJ/cm$^2$.

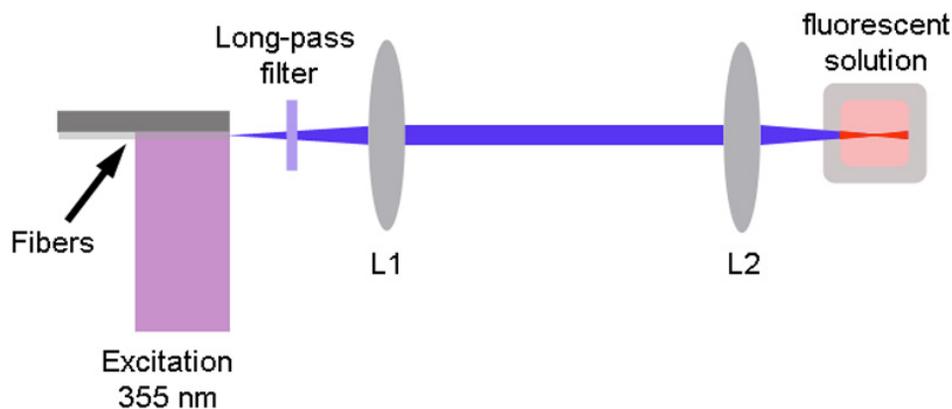

**Figure S6.** Experimental scheme for solution excitation by ASE from UV-emitting fibers. A long-pass filter removes the residual laser radiation, whereas the emission from fibers is collimated and refocused by a double lens system ($L_1$, $L_2$).